\documentclass[pra,aps,twocolumn,showpacs,superscriptaddress,nofootinbib]{revtex4}

\usepackage{graphicx}
\usepackage{amssymb}
\usepackage{amsmath}
\usepackage{color}

\newtheorem{thrm}{Theorem}

\newcommand{\qed}{\hfill $\blacksquare$}

\newtheorem{defn}{Definition}

\newcommand{\ket}[1]{|#1 \rangle}
\newcommand{\bra}[1]{\langle #1 |}

\newcommand{\tr}{\mathrm{Tr}}

\newcommand{\trho}{\tilde{\rho}}

\begin{scriptsize}
\begin{footnotesize}
\begin{small}
\begin{normalsize}
\end{normalsize}
\end{small}
\end{footnotesize}
\end{scriptsize}

\begin{document}

\title{Operational Significance of Discord Consumption: Theory and Experiment}

\author{Mile Gu}
\affiliation{Centre for Quantum Technologies, National University of Singapore, Singapore}

\author{Helen M. Chrzanowski}
\affiliation{Centre for Quantum Computation and Communication Technology, Department of Quantum Science, The Australian National University, Canberra, ACT 0200, Australia}

\author{Syed M. Assad}
\affiliation{Centre for Quantum Computation and Communication Technology, Department of Quantum Science, The Australian National University, Canberra, ACT 0200, Australia}

\author{Thomas Symul}
\affiliation{Centre for Quantum Computation and Communication Technology, Department of Quantum Science, The Australian National University, Canberra, ACT 0200, Australia}

\author{Kavan Modi}
\affiliation{Atomic and Laser Physics, Clarendon Laboratory, University of
Oxford, United Kingdom}
\affiliation{Centre for Quantum Technologies, National University of Singapore, Singapore}

\author{Timothy C. Ralph}
\affiliation{Centre for Quantum Computation and Communication Technology, Department of Physics, University of Queensland, St Lucia 4072, Australia}

\author{Vlatko Vedral}
\affiliation{Centre for Quantum Technologies, National University of Singapore, Singapore}
\affiliation{Department of Physics, National University of Singapore, Singapore}
\affiliation{Atomic and Laser Physics, Clarendon Laboratory, University of
Oxford, United Kingdom}

\author{Ping Koy Lam}
\affiliation{Centre for Quantum Computation and Communication Technology, Department of Quantum Science, The Australian National University, Canberra, ACT 0200, Australia}

\date{\today}

\begin{abstract}
Coherent interactions that generate negligible entanglement can still exhibit unique quantum behaviour. This observation has motivated a search beyond entanglement for a complete description of all quantum correlations. Quantum discord is a promising candidate~\cite{Henderson01, Zurek01a}. Here, we demonstrate that under certain measurement constraints, discord between bipartite systems can be consumed to encode information that can only be accessed by coherent quantum interactions. The inability to access this information by any other means allows us to use discord to directly quantify this `quantum advantage'. We experimentally encode information within the discordant correlations of two separable Gaussian states. The amount of extra information recovered by coherent interaction is quantified and directly linked with the discord consumed during encoding. No entanglement exists at any point of this experiment. Thus we introduce and demonstrate an operational method to use discord as a physical resource.
\end{abstract}

\maketitle

Correlations lie at the heart of our capacity to manipulate information. The fewer the constraints on the correlations we can exploit, the greater our capacity to manipulate information in ways we desire. The rapid development of quantum information science is a testament to this observation. Quantum systems may be so correlated that they are `entangled', such that each of its subsystems possesses no local reality. Exploitation of such uniquely quantum correlations has led to many remarkable protocols that would otherwise be either impossible or infeasible~\cite{Bennett92c, Shor97a, Ekert91a, Grover97a}.

However, the absence of entanglement does not eliminate all signatures of quantum behaviour~\cite{Bennett84a}. Coherent quantum interactions (i.e., quantum two-body operations) between separable systems that result in negligible entanglement could still lead to exponential speed-ups in computation~\cite{PhysRevLett.81.5672, vidal07, datta08, Lanyon08}, or the extraction of otherwise inaccessible information~\cite{Bennet99c}. The potential presence of discord~\cite{Henderson01, Zurek01a} within such protocols motivated speculation that discord could prove a better quantifier of the `quantum resource' that coherent interactions exploit to deliver a `quantum advantage'~\cite{arXiv:quant-ph/0110029, datta08, vedral10}. Discord has thus captured a great deal of attention, as evidenced by studies of its role in open dynamics~\cite{arXiv:quant-ph/0703022}, cloning of correlations~\cite{arXiv:0707.0848, PhysRevA.82.012338}, scaling laws in many-body physics~\cite{arXiv:1012.4270}, and quantum correlations within continuous variable systems~\cite{Giorda10, Adesso10}.

Discord is related to fundamental processes of physics as well as information-theoretic protocols. The original motivations for discord were to understand the correlation between a quantum system and classical apparatus~\cite{Zurek01a} and the division between quantum and classical correlation~\cite{Henderson01}. A similar quantity called deficit
was employed to study the thermodynamic work extraction and Maxwell's demon \cite{oppenheim, zurek03, arXiv:1002.4913}. The existence of discord puts constraint on broadcasting or cloning of states under local operation\cite{arXiv:0707.0848, PhysRevA.82.012338}. Discord is equal to the amount classical correlation that can be unlocked in quantum-classical states \cite{arXiv:0811.4003, Boixo11}. Discord between two parties is related to the resource for quantum state merging with another party \cite{kavan11,madhok11}. Here we introduce and experimentally verify a protocol for which the consumption of discord is directly related to the advantage of coherent interactions.

Alice encodes information within one arm of a bipartite quantum state $\rho_{AB}$. Bob is tasked to retrieve the encoded data. We compute Bob's optimal performance when he is restricted to performing a single local measurement on each bipartition (i.e., Bob can make a local measurement first on $A$, then $B$, or vice versa). We compare this to the case where Bob can, in addition, coherently interact the bipartitions, which allows him to effectively measure in an arbitrary joint basis of $A$ and $B$. We show that coherent two-body interactions are advantageous if and only if $\rho_{AB}$ contains discord and that the amount of discord Alice consumes during encoding bounds exactly this advantage. This indicates discord quantifies a resource that can be consumed to give coherent interactions an operationally meaningful advantage. During the preparation of this manuscript, discord was shown to quantify the performance loss in a broad class of quantum communication protocols, when one of the parties suffers decoherence \cite{arXiv:1107.0994}. When applied to dense coding, their results may be interpreted as an instance of our protocol, where Alice's encoding consumes all the discord in $\rho_{AB}$ and Bob is constrained to measuring $B$ before $A$.

We experimentally implement such a protocol in the continuous variables regime. Our results show that even in the presence of experimental imperfections, coherent quantum interactions can harness discord to extract information many standard deviations beyond what is possible otherwise. Furthermore, our experimental implementation contains no entanglement at any point. This confirms that discord alone, without entanglement, is a sufficient resource for coherent interactions to deliver an observable advantage for the task at hand.

\section{Theory}
Discord is defined by the difference between two different measures of correlations within a bipartite quantum system in state $\rho_{AB}$, which we denote by $I(A,B)$ and $J(A|B)$. $I(A,B) = S(\rho_A) + S(\rho_B) - S(\rho_{AB})$, represents the total correlations between the two subsystems, where $\rho_{A}$ and $\rho_{B}$ are states of the respective subsystems, and $S(\cdot)$ is the von Neumann entropy. Meanwhile $J(A|B) = S(\rho_A) - \max_{\{\Pi_b\}\in\mathcal{M}}\sum p_b S(\rho_{A|b})$ represents the contribution of classical correlations; and is defined by the reduction in the entropy of $A$ after a measurement on $B$, when maximized over a class of measurements $\mathcal{M}$. Here, $p_b$ is the probability of getting measurement outcome $b$, leaving $A$ in the conditional state $\rho_{A|b}$, and $\mathcal{M}$ represents some class of viable measurements. We will first assume that $\mathcal{M}$ is the set of all positive operator value measurements (POVMs), then discuss other scenarios. The discrepancy $\delta(A|B) = I(A,B) - J(A|B)$ defines the discord. If we instead consider measurements on $A$, we have $\delta(B|A) = I(A,B) - J(B|A)$. In general $\delta(A|B) \neq \delta(B|A)$.

Alice prepares some correlated resource $\rho_{AB}$ on a bipartite quantum system. She labels the bipartitions $A$ and $B$ such that $\delta(A|B) \leq \delta(B|A)$ and gives $B$ to Bob. Alice then privately encodes a random variable $\mathbf{K}$ with probability $P(K = k) = p_k$ onto her subsystem by application of a corresponding unitary operator $U_k$. The preparation and encoding scheme is publicly announced. To anyone oblivious to which $U_k$ was applied, this results in the encoded state
\begin{equation}
\trho_{AB} = \sum_k p_k U_k \rho_{AB} U^\dag_k,
\end{equation}
with corresponding discord $\tilde{\delta}(A|B)$. Thus $\Delta \delta(A|B) = \delta(A|B) - \tilde{\delta}(A|B)$
represents the amount of discord consumed to during encoding. 

Alice then gives system $A$ to Bob and challenges him to retrieve the best possible estimate of $\mathbf{K}$ from $\trho_{AB}$. To meet this challenge, Bob would need to apply some decoding protocol, defined by a computational process on $\trho_{AB}$, that outputs a classical variable $\mathbf{K}_o$. The performance of his chosen protocol is then determined by the classical mutual information $I(\mathbf{K}_o,\mathbf{K})$.

Let $I_c$ be Bob's best possible performance when he is restricted to a single local measurement on each of $A$ and $B$ and classical post-processing. Let $I_q$ be Bob's maximum performance when he can, in addition, implement arbitrary quantum operations between $A$ and $B$, and thus effectively optimize his performance over all possible basis measurements of the joint system $AB$. The difference $\Delta I  = I_q - I_c$ defines the extra \emph{quantum advantage} that coherent interactions can potentially deliver. In the supplementary materials, we prove that
\begin{equation}\label{eqn:symmetric_relation}
\Delta\delta(A|B)  - \tilde{J}(A|B) \leq \Delta I \leq \Delta\delta(A|B),
\end{equation}
where $\tilde{J}(A|B)$ represents the amount of classical correlations \emph{after} encoding (i.e., in $\trho_{AB}$). This equation allows as to answer the question: \emph{when are coherent interactions advantageous for Bob}?

Eqn.~(\ref{eqn:symmetric_relation}) indicates that discord indeed quantifies a resource coherent quantum interactions harness to deliver an otherwise impossible advantage. $\Delta I \leq \Delta\delta(A|B)$ implies this advantage exists only when discord is consumed during the encoding process. Furthermore, the magnitude of this advantage is bounded above by the amount of discord consumed. In particular, should Alice and Bob possess no discord to begin with, i.e., $\delta(A|B) = 0$, there exists no discord to consume (i.e. $\Delta\delta(A|B) = 0$), and in consequence coherent interactions will be of no help to Bob regardless of Alice's choice of encoding.

Meanwhile the lower bound $\Delta\delta(A|B)  - \tilde{J}(A|B) \leq \Delta I$ indicates that quantum advantage is guaranteed for any encoding such that the discord consumed is strictly greater the classical correlations after encoding. This is possible for any discorded $\rho_{AB}$, since there exists since there exists maximal encodings, such that $\tilde{J} = \tilde{\delta} = 0$ for any $\rho_{AB}$ (see supplementary materials). In this scenario, all available discord initially available is consumed, the lower and upper bounds converge, and eqn. (\ref{eqn:symmetric_relation}) reduces to
\begin{equation}\label{eqn:symmetric_relation_limit}
\Delta I = \Delta\delta(A|B) = \delta(A|B).
\end{equation}
Discord therefore quantifies exactly a resource that coherent interactions can exploit. An example of maximal encoding on two qubits are the Pauli operators $\{I,\sigma_x,\sigma_z,\sigma_x\sigma_z\}$ chosen with equal probability. The special case where this encoding is applied to a singlet state coincides with dense coding~\cite{Bennett92c}. Coherent interactions allow Bob to extract one extra bit of knowledge about which of the four unitary transformations was applied by Alice. This equals the discord consumed when we encode onto the singlet state.

The operational significance of discord beyond entanglement is highlighted when we repeat the above protocol on a separable discordant resource. For example, take $\rho_{AB} = \sum_{i = \{x,y,z\}}(\ket{0}_i\ket{0}_i\bra{0}_i\bra{0}_i + \ket{1}_i\ket{1}_i\bra{1}_i\bra{1}_i)$, where $\ket{0}_i$ and $\ket{1}_i$ represent the computational basis states with respect to $\sigma_i$. This resource is clearly separable, and yet possesses a discord of $\frac{1}{3}$. Therefore coherent processing can harness the discord within this resource to extract $\frac{1}{3}$ extra bits of information despite the absence of entanglement.

\begin{figure*}
\begin{center}
\includegraphics[width=20cm]{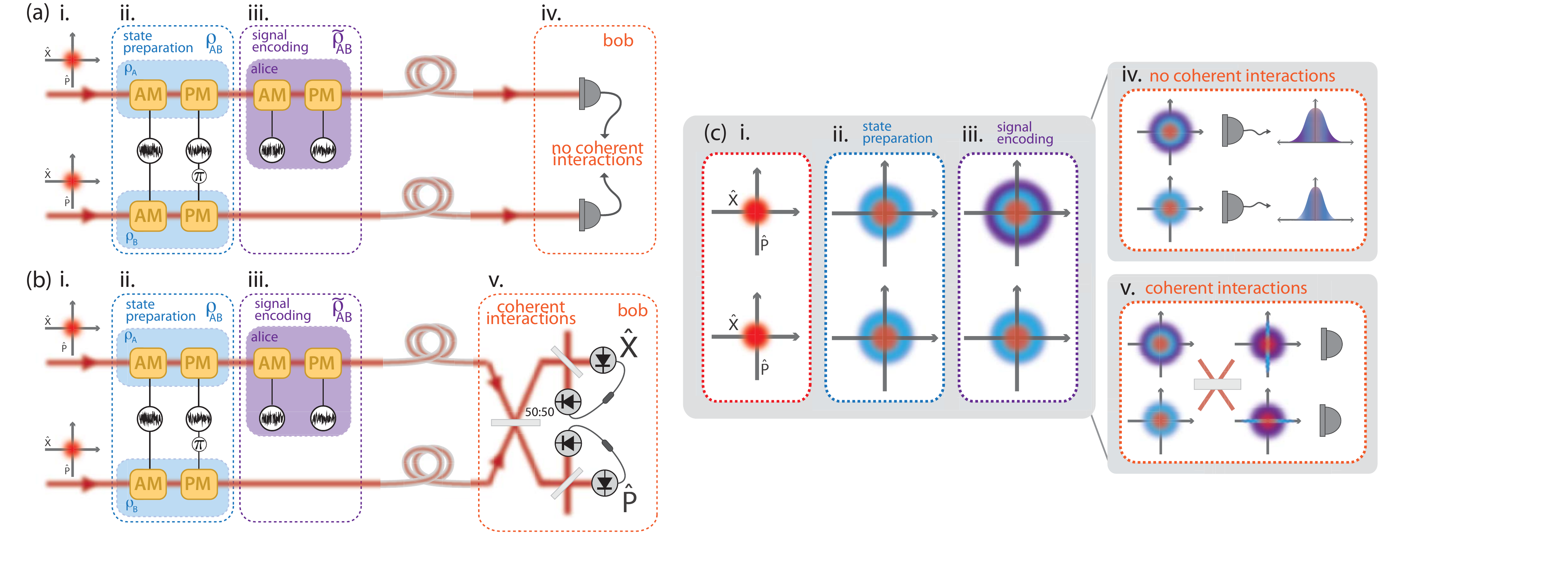}
\caption{
{\bf Experimental Setup of the protocol when Bob  (a) is limited to a single measurement on each bipartition and (b) can in addition, coherently interact the bipartitions. (c) Phase space representations of the bipartite state encoding and processing.} (i) A laser provides coherent light that is encoded using modulation of the sideband frequencies between 3.2-3.8 MHz around the carrier. (ii) The bipartite state $\rho_{AB}$ is prepared by correlated (anti-correlated) displacement of two coherent vacuum states in the amplitude (phase) quadrature with Gaussian distributed noise. This state is shared between Alice and Bob. This is experimentally realized through electro-optic modulation (EOM) of the phase and amplitude quadrature using independent classical Gaussian noise. (iii) Alice then encodes independent signals $\mathbf{X}_s$ and $\mathbf{Y}_s$ on the phase and amplitude quadrature of her subsystem using EOM and subsequently transmits her state to Bob. We compare Bob's capacity to extract information in two different scenarios. The theoretical limit to Bob's performance when Bob makes a local measurement of $A$ and $B$ (iv) and the experimental observed performance when Bob uses a particular protocol involving coherent interference to enhance his knowledge of Alice's encoding. Whilst in (iv) Bob can only harness the classical correlations in the bipartite system to gain knowledge regarding Alice's encoding, in (v), he can exploit quantum and classical correlations through coherent interaction to enhance his knowledge of Alice's encoding. }
\label{fig1}
\end{center}
\end{figure*}

So far, we have assumed in our definition of $J(A|B)$ that Bob may choose any POVM to gain information about Alice's encoding. Depending on context, the class of measurements, $\mathcal{M}$, that we optimize over to obtain $J(A|B)$ is sometimes restricted to only projective measurements~\cite{Zurek01a}; and in the case of continuous variables, sometimes Gaussian measurements~\cite{Giorda10,Adesso10}. Our results can be adapted to such variants, where $\Delta I$ now bounds the extra advantage gained by Bob if he can implement arbitrary coherent interactions, when restricted to measurements within $\mathcal{M}$.

\section{Experiment}
The resource based view of discord allows for an explicit class of experiments that test for the discord-induced advantage of coherent interactions. We can generate a non-entangled discordant state $\rho_{AB}$ on two spatially separate subsystems by local operations and classical communication (LOCC), thus ensuring that any observed advantage of coherent interactions is due solely to discord.  Discord is consumed by encoding a classical variable $\mathbf{K}$ on one subsystem. We attempt to estimate $\mathbf{K}$ by measurement in a basis the requires coherent interaction of $A$ and $B$. Provided our estimate $\mathbf{K}_{o}$ satisfies $I(\mathbf{K}_{o}, \mathbf{K}) - I_c = \Delta I^{\rm exp} > 0$, we can then guarantee that our interactions are indeed coherent, and allow extraction of $\Delta I^{\rm exp}$ bits beyond the incoherent limit. This proposal is not restricted to any particular bipartite system, or any particular encoding. It is therefore applicable to a wide range of experimental platforms.

In continuous variable (CV) Gaussian optics, quadrature measurements can be performed at high fidelity, while the coherent interaction of separate optical modes requires only a beamsplitter \cite{Brau01a,Fiura01a}. These features make the CV regime ideal for engineering a coherent quantum interaction at the precision necessary to exceed the incoherent limit. Recent work has developed a theoretical framework for discord in Gaussian systems where $\mathcal{M}$ is the set of all Gaussian measurements~\cite{Giorda10,Adesso10}. Our experimental setup is shown in Fig.~\ref{fig1}. Alice and Bob share a bipartite state $\rho_{AB}$ that is prepared by displacing two independent vacuum states (see Fig.~\ref{fig1}.i.) with correlated and anti-correlated Gaussian distributed noise of variance $V$ in the phase and amplitude quadratures (see Fig.~\ref{fig1}.ii.) respectively. As this procedure involves only LOCC, the resulting bipartite state is clearly separable. However, the non-commutation of amplitude and phase quadrature operators, $X$ and $Y$, result in a Gaussian discord of
\begin{equation}
\delta({A|B}) = g\left(V+1\right) - 2g \left(\sqrt{2V + 1}\right) + g\left(1 + \frac{2V}{2+V}\right)
\end{equation}
between the bipartitions \cite{Giorda10,Adesso10}, where $g(x) = x_+ \log x_+ - x_- \log x_-$, and $x_\pm = \frac{1}{2}(x \pm 1)$. We refer to $V$ as the discording noise.
\begin{figure}
\begin{center}
\includegraphics[width=\columnwidth]{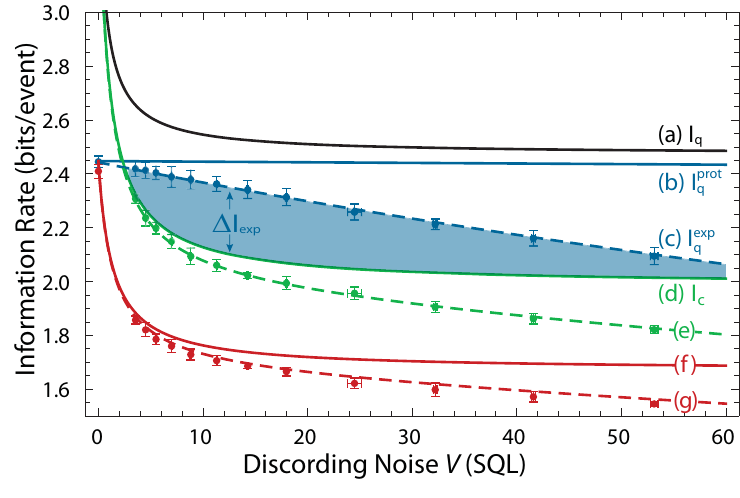}
\caption{Plot of Bob's knowledge of the encoded signal for bipartite resource states with varying discording noise and fixed encoding variance $V_s$. (a) represents the amount of information Bob can theoretical gain should he be capable of coherent interactions. For our proposed implementation, this maximum is reduced to (b). Experimentally, Bob's knowledge about the encoded signal is represented by the blue data points. The line (c) models these observations by taking experimental imperfections into account. Despite these imperfections, Bob is still able to gain more information than the incoherent limit (d). The shaded region highlights this quantum advantage. This advantage is more apparent if we compare Bob's performance to the reduced incoherent limit when experimental imperfections are accounted for (e). We can also compare these rates to a practical decoding scheme for Bob when limited to a single measurement on each optimal mode (f) and its imperfect experimental realization (g) (see supplementary material).}
\label{fig2}
\end{center}
\end{figure}
\begin{figure}
\begin{center}
\includegraphics[width=\columnwidth]{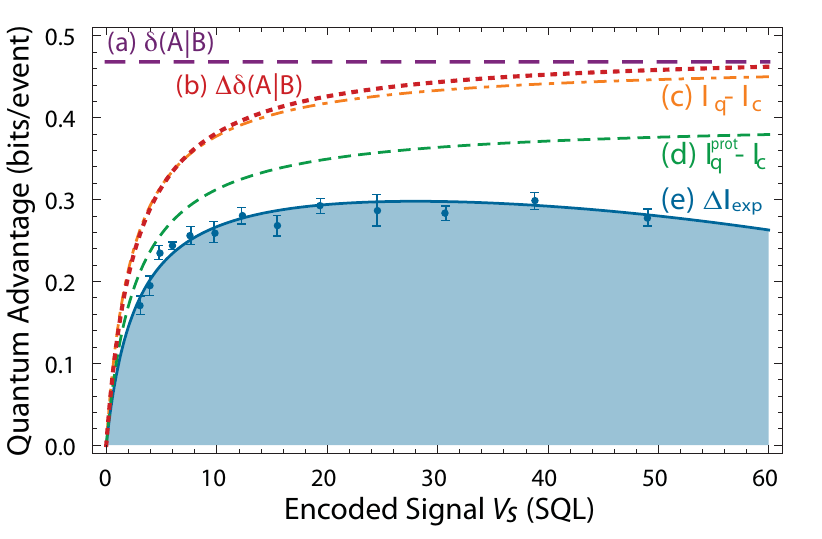}
\caption{Plot of quantum advantage for a fixed resource state (with $V = 10.0 \pm 0.1 $) with varying strength of the the encoded signal, $V_s$. (a) represents the maximum available amount of discord in the original resource $\rho_{AB}$, of which we progressively consume more of as we increase $V_s$ (b). This bounds the maximum possible quantum advantage, assuming Bob can perform an ideal decoding protocol that saturates the Holevo limit (c). In the limit of large $V_s$, the encoding becomes maximal, and this tends to the discord of the original resource (a). The actual advantage that can be harnessed by our proposed protocol is represented by (d). In practice, experimental imperfections reduce the experimentally measured advantage to (e).}
\label{fig3}
\end{center}
\end{figure}
To quantify the discord induced quantum advantage, we encode separate Gaussian signals $\mathbf{X}_s$ and $\mathbf{Y}_s$ of equal variance $V_s$ in the phase and amplitude quadrature of her beam (see Fig.~\ref{fig1}.iii.).  The encoded state $\trho_{AB}$ is completely specified by the covariance matrix
\begin{equation}\label{eqn:pmatrix}
\mathbf{\sigma}(\trho_{AB}) = \begin{pmatrix}
V+1 & 0 & V & 0 \\
0 & V+1 & 0 & -V \\
V & 0 & V+V_s+1 & 0 \\
0 & -V & 0 & V+V_s+ 1 \\
\end{pmatrix}.
\end{equation}
The amount of discord consumed during encoding, $\Delta \delta(A|B)$, grows monotonically with $V_s$ (Fig. \ref{fig3}.b). In the limit of large $V_s$, the encoding becomes approximately maximal.

Bob is then required to extract as much information regarding the encoded signal $(\mathbf{X}_s \mathbf{Y}_s)$. Should Bob be limited to a single measurement on each beam, his knowledge of $({\bf X}_s, {\bf Y}_s)$ will be bounded by
\begin{equation}\label{eqn:Ic}
I_c = g\left(1 + \frac{2V}{V+2} + V_s\right) - g\left(1 + \frac{2V}{2+V}\right).
\end{equation}
In contrast, if Bob has the capacity to implement arbitrary coherent interactions, he can theoretically achieve a performance of
\begin{equation}\label{eqn:cv_q_ideal}
I_q = g(\mu_+) + g(\mu_-) - 2 g(\sqrt{2V + 1}),
\end{equation}
where $\mu_{\pm}$ are the symplectic eigenvalues of $\mathbf{\sigma}(\trho_{AB})$ (see supplementary materials). In the limit of large $V_s$, the encoding consumes all the discord in $\rho_{AB}$, and $\Delta I = I_q - I_c$ coincides with the discord $\delta(A|B)$ within the original resource (Fig. \ref{fig3}.c).

To experimentally measure $\Delta I$, we implement a protocol that exploits coherent quantum interactions to retrieve information regarding $(\mathbf{X}_s, \mathbf{Y}_s)$ beyond the incoherent limit, $I_c$. For each trial of the experiment with a different value of $V$ and $V_s$, we first characterize the co-variance matrix of the encoded bipartite system $\trho_{AB}$. This then allows direct inference of the theoretical incoherent limit $I_c$ from the Holevo limit. To obtain $I_q^{\rm exp}$ we interfere the two beams in phase on a 50:50 beam-splitter (Fig 1.v). The outputs are then measured via balanced homodyne detection, which result in observables $(\mathbf{X}_o,\mathbf{Y}_o)$. We characterize the information Bob can retrieve about the encoded signal by measurement of the mutual information
\begin{equation}
I_q^{\rm exp} = I[(\mathbf{X}_o,\mathbf{Y}_o),(\mathbf{X}_s,\mathbf{Y}_s)].
\end{equation}
$I_q^{\rm exp}$ is directly measured from the resulting signal to noise ratios between $({\bf X}_s,{\bf Y}_s)$ and $({\bf X}_o,{\bf Y}_o)$. This procedure allows experimental observation of $\Delta I^{\rm exp}$. Provided $\Delta I^{\rm exp} \equiv I_q^{\rm exp} - I_c > 0$, Bob has extracted some information that can only be accessed by coherent interactions. $\Delta I^{\rm exp}$ defines the amount of discord-induced quantum advantage we experimentally observe. For a perfect realisation of this protocol, one can verify that $I_q^{\rm exp} \rightarrow I_q^{\rm prot} \equiv \log\left(1 + V_s/2\right)$. While this protocol does not saturate Eqn. \ref{eqn:cv_q_ideal}, it is ideal in the limit of large discording noise, i.e., $\lim_{V \rightarrow \infty}I^{\rm prot}_q \rightarrow I_q$. In addition, since the entire protocol involves only passive linear optics and non-squeezed sources, there exists no entanglement at any point. 

Figure~\ref{fig2} details our experimental results for bipartite resources with varying discording noise and a fixed signal variance (normalized to the standard quantum limit) of $9.10\pm 0.05$. The blue data points represent the observed values of $I^{\rm exp}_q$. No corrections for experimental imperfections are made. Provided the discording noise is sufficiently large, such that the original resource has a significant amount of discord, $I^{\rm exp}_q$ clearly exceeds the incoherent limit $I_c$ (Fig.~2.d). There is a noticeable deviation between the amount of information we experimentally extract, and the theoretical prediction $I_q^{\rm prot}$ of the idealized protocol (Fig 2.b). This is due to experimental imperfections, which include losses, limited suppression of parasitic phase and amplitude modulations, asymmetric modulation depth, and non-ideal 50:50 beam splitter at the interference stage. When these errors are taken into account, theory and observation agree within experimental error (Fig.~2.c). The shaded region then gives the observed advantage $\Delta I^{\rm exp}$ of coherent interactions.

Figure~\ref{fig3} gives the experimentally observed quantum advantage for $\rho_{AB}$ with varying strength of the encoded signal, $V_s$, when the discording noise is fixed at $V = 10.0 \pm 0.1$ normalised to shot noise. The amount of discord within the initial resource is fixed at $\delta(A|B)$ (Fig 3.a). This quantifies the correlations shared by Alice and Bob that can be potentially harnessed to exhibit quantum advantage. As we increase the strength of the encoded signal, progressively more of this initial resource is consumed (Fig 3.b), and thus bounds how much of this discord can be potentially harnessed (Fig.~3.c). In an ideal version of the decoding protocol, the advantage would increase monotonically with the signal strength (Fig.~3.d). With our imperfect experimental setup, there is initially an increase in the observed quantum advantage for increasing signal. However, there exists a saturation point around $V_s \sim 20$, beyond which the extra theoretical gain from increased signal strength is offset by the extra experimental imperfections in encoding. This is attributed to the nonlinear response of the electro-optic modulators. When we include these imperfections within our theoretical model, observations and theory agree (Fig.~3.d).

\section{Conclusion and Discussion}
In this article, we have demonstrated that coherent interactions can harness discord to complete a task that is otherwise impossible. Experimental implementation of this task demonstrates that this advantage can be directly observed, even in the absence of entanglement. Since the capacity to coherently interact quantum systems is essential to quantum processing, our results indicate that some of the advantages quantum processing pertains over its classical counterpart can be attributed to its potential to harness discord.

The relation between the advantage of coherent interactions and non-classical correlations has also been studied within related paradigms. The thermodynamic variant of discord, for example, characterizes the advantage of coherent interactions in energy extraction from a given quantum state \cite{oppenheim, zurek03, arXiv:1002.4913}. Our protocol gives similar interpretation for the standard notion of discord in terms of information extraction. Meanwhile, the consideration of discord in the `mother protocol' shows that discord quantifies the advantage of coherent interactions in undoing entangling operations between one of the parties and an ancillary system for a general class of memory-less, two party, communication protocols \cite{arXiv:1107.0994}. By adopting the framework of the mother protocol, we may be able to demonstrate that the consumption of discord is a general resource that induces advantage in the coherent interaction between the two relevant parties.

In addition, the capacity for coherent interactions that harness discord to improve performance allows potential reinterpretion of many existing protocols. If we regard our proposal as an attempt for Alice to communicate the contents of $\mathbf{K}$ to Bob via a pre-shared resource $\rho_{AB}$, the protocol resembles a quantum one-time pad with a generic resource \cite{schumacher06}. Discord now plays a role in measuring the amount of extra information coherent interactions can unlock. In the special case where $A$ and $B$ are entangled, this protocol corresponds to dense coding, where the additional gain in communication rates is made possible by coherent interactions that decode information within the discordant correlations. Meanwhile, if we regard the task of trace estimation in `deterministic quantum computing with one qubit' (DQC1)~\cite{PhysRevLett.81.5672} as Bob's attempt to extract information Alice has encoded within the trace of a given unitary, our protocol may shed light on where the power in DQC1 originates. These connections are worth further investigation, and may not only lead to additional insight on the role of discord within a diverse range of applications, but also indicate whether we are already harnessing discord in many existing proposals without realizing it.

\section*{Acknowledgements}
We thank C. Weedbrook, J. Thompson, N. Walk and H. Wiseman for helpful discussions. The research is supported by the National Research Foundation and Ministry of Education in Singapore (MG, KM and VV), the John Templeton Foundation (KM and VV), and the Australian Research Council Centre of Excellence for Quantum Computation and Communication Technology, project number CE11E0096 (HC, SMA, TS, TCR and PKL).


\begin{appendix}

\section*{Supplementary Materials}
\section{Proof that Discord is a quantifier of quantum advantage}
In this section, we explicitly prove that
\begin{equation}
\Delta \delta(A|B) - \tilde{J}(A,B) \leq \Delta I \leq \Delta \delta(A|B),
\end{equation}
which is equivalent to the statement $\delta(A|B) - \tilde{I}(A,B) \leq \Delta I \leq \delta(A|B) - \tilde{\delta}(A|B)$, where $\tilde{I}(A,B)$ denotes the mutual information of $\trho_{AB}$. To do this, we make use of the Holevo information. Let $\mathbf{K}$ be a random variable that takes on value $k$ with probability $p_k$. If each $k$ is encoded in a quantum state with density operator $\rho_k$, then the maximum amount of information that may later be extracted about $\mathbf{K}$ is given by
\begin{equation}\label{eq:helevo}
S\left(\sum_k p_k \rho_k\right) - \sum_k p_k S(\rho_k).
\end{equation}
when there are no constraints on what quantum operations are allowed.

To evaluate $\Delta I$, we first introduce an additional scenario where Bob has no access to system $B$, and attempts to the find the best estimate of $\mathbf{K}$ using only measurements on system $A$. Let $I_0$ be Bob's maximum performance in this scenario. Recall that after encoding, the bipartite state between Alice and Bob is given by
\begin{equation}
\trho_{AB} = \sum_k p_k U_k \rho_{AB} U^\dag_k
\end{equation}

Since Bob has no access to $B$, we can trace over system $B$. Noting that $U_k$ acts only on system $A$ and is thus preserved under the partial trace, this results in codewords $U_k \rho_A U_k^\dag$, which give Bob
\begin{equation}\label{eq:i0}
I_0 = S(\trho_A) - S(\rho_A)
\end{equation}
bits of accessible information by application of Eqn. (\ref{eq:helevo}). Here, $\trho_A = \tr_B(\sum_k p_k U_k \rho_{AB} U_k^\dag) = \sum_k p_k U_k \rho_A U_k^\dag$. This case can be considered the control, i.e., the amount of information accessible to Bob when he cannot access any of the correlations between $A$ and $B$.

We now compute $I_q$, the maximum extra information available to Bob when he can implement arbitrary interaction between $A$ and $B$. In this case, we have codewords $\rho_k = U_k \rho_{AB} U_k^\dag$, such that $S(\rho_k) = S(\rho_{AB})$. This results in a Holevo information of $I_q = S(\trho_{AB}) - S(\rho_{AB})$. Therefore, the extra information Bob gains over the control case is
\begin{equation}\label{eq:iq}
\Delta_q \equiv I_q - I_0 = S(\trho_{AB}) - S(\rho_{AB}) - S(\trho_A) + S(\rho_A) = I(A,B) - \tilde{I}(A,B).
\end{equation}
$I$ and $\tilde{I}$ respectively represent the total correlations between $A$ and $B$ before and after encoding. Thus, the advantage of being able to implement arbitrary two-body interactions over having no way to make use of system $B$ coincides with the total amount of correlations consumed during the encoding process.

Similarly, we compute $I_c$, the maximum amount of information available to Bob by a single local measurement on each bipartition. We note that this constraint is equivalent to local operations and one-way communication, since multiple rounds of two way communication does not help Bob if does not measure a single partition more than once. In this scenario, the best Bob can do is to first measure either $A$ or $B$ in some basis $\{\Pi_b\}$, and make use of the classical output to improve his estimate of $\mathbf{K}$.

Consider first a measurement on $B$. Let Bob's resulting performance be $\overleftarrow{I}_c$. The state after measurement is $\rho_{A|b} = \tr_B(\rho_{AB} \Pi_b)/q_b$ with probability $q_b$, where $q_b = \tr(\rho_{AB} \Pi_b)$. Thus, Alice has effectively encoded $\mathbf{K}$ onto codewords $U_k \rho_{A|b} U_k^\dag$. This results in $S(\sum_k p_k U_k \rho_{A|b} U_k^\dag) - S(\rho_{A|b})$ bits of information accessible about $\mathbf{K}$ with probability $q_b$.  To obtain the upper bound on how much information accessible to Bob, we maximize the expected value of the above subject to all possible measurements Bob could have made, thus
\begin{equation}\label{eq:ic}
\overleftarrow{I}_c = \sup_{\{\Pi_b\}} \left( \sum_b q_b S (\trho_{A|b}) - \sum_b q_b S(\rho_{A|b}) \right).
\end{equation}
Here, we have used the fact that Alice's application of $U_k$ on system $A$, and Bob's measurement of system $B$ act on different Hilbert spaces, and thus commute.

Now consider the case where Bob first measures $A$. We partition to total amount of information Bob can gain, $\overrightarrow{I}_c$ into two components; the component $\overrightarrow{I}^{(A)}_c$, that he gains directly from his measurement of system $A$; and $\overrightarrow{I}^{(B)}_c$, that he gains from the resulting collapsed quantum state on system $B$. Clearly $\overrightarrow{I}^{(A)}_c = I_0$, since Bob has not yet measured $A$.

To bound $\overrightarrow{I}^{(B)}_c$, note that measurement of $\rho_k = U_k \rho_{AB} U^\dag_k$ on system $A$ in a basis $\{\Pi_a\}$ is equivalent to measurement of $\rho_{AB}$ in a rotated basis $\{U_k^\dag \Pi_a U_k\}$. Thus, for each possible encoding $k$, the entropy of $B$ after measurement is bounded below by $\sum_a \inf_{\{\Pi_a\}}S(\rho_{B|a})$. Therefore, the Holevo bound gives $\overrightarrow{I}^{(B)}_c \leq S(\rho_B) - \inf_{\{\Pi_a\}}\sum_a S(\rho_{B|a})$, and thus
\begin{equation}\label{eq:icreverse}
\overrightarrow{I}_c \leq \overrightarrow{I}^{(A)}_c + \overrightarrow{I}^{(B)}_c \leq I_0 + S(\rho_B) -
\inf_{\{\Pi_a\}} \sum_a  S(\rho_{B|a}).
\end{equation}
The optimal amount of information Bob can extract without coherent interactions is thus the maximal of $\overleftarrow{I}_c$ and $\overrightarrow{I}_c$, i.e.,
$I_c = \max\{\overrightarrow{I}_c,\overleftarrow{I}_c\}$. Noting that, $\Delta_c = I_c - I_0 = \max\{\overrightarrow{\Delta}_c,\overleftarrow{\Delta}_c\}$, where $\overrightarrow{\Delta}_c = \overrightarrow{I}_c - I_0$ and $\overleftarrow{\Delta}_c = \overleftarrow{I}_c - I_0$, we first evaluate $\overrightarrow{\Delta}_c$ and $\overleftarrow{\Delta}_c$ separately.

Substraction of Eq. (\ref{eq:i0}) from Eq. (\ref{eq:ic}) gives
\begin{equation}\label{deltaceq}
\overleftarrow{\Delta}_c = S(\rho_A) + \sup_{\{\Pi_b\}}\left(\sum_b q_b S(\trho_{A|b}) - S(\trho_{A}) - \sum_b q_b S(\rho_{A|b})\right).
\end{equation}
Noting that $S[\trho_{A|b}] \leq S[\trho_{A}]$ since entropy can never increase under conditioning, we immediately find
\begin{equation}
\overleftarrow{\Delta}_c \leq S(\rho_A) - \inf_{\{\Pi_b\}}\sum_b q_b S(\rho_{A|b}) = J(A|B).
\end{equation}
Also, rearranging Eq. \ref{deltaceq} gives
\begin{align}\nonumber
\overleftarrow{\Delta}_c = & \sup_{\{\Pi_b\}} \Bigg[S(\rho_A) - \sum_b q_b S(\rho_{A|b}) \\ \nonumber
- & \left(S(\trho_{A})- \sum_b q_b S(\trho_{A|b})\right)\Bigg],\\ \nonumber
\geq & \sup_{\{\Pi_b\}} \left[S(\rho_A) - \sum_b q_b S(\rho_{A|b})\right] \\ \nonumber
&- \sup_{\{\Pi_b\}}\left[S(\trho_{A})- \sum_b q_b S(\trho_{A|b})\right],\\ \nonumber
= & S(\rho_A) - \inf_{\{\Pi_b\}} \sum_b q_b S(\rho_{A|b})
  \\ \nonumber
& - \left(S(\trho_{A})-\inf_{\{\Pi_b\}}\sum_b q_b S(\trho_{A|b})\right)\\
= & J(A|B) - \tilde{J}(A|B),
\end{align}
where $\tilde{J}(A|B) = S(\trho_A) - \inf_{\{\Pi_b\}}\sum_b S(\trho_{A|b})$ denote the classical correlations of $\trho_{AB}$.

Therefore
\begin{equation}
J(A|B) - \tilde{J}(A|B) \leq \overleftarrow{\Delta}_c \leq J(A|B),
\end{equation}
Meanwhile, subtraction Eq. (\ref{eq:i0}) from Eq. (\ref{eq:icreverse}) gives $\overrightarrow{\Delta}_c = \overrightarrow{I}_c - I_0 \leq \overrightarrow{J}$,
therefore
\begin{equation}
J(A|B) - \tilde{J}(A|B) \leq \Delta_c \leq \max \{J(A|B),J(B|A)\}.
\end{equation}
Subtraction of this equation from Eqn.~$(\ref{eq:iq})$ immediately bounds the extra performance of coherent processing over its incoherent counterpart.
\begin{align}\label{eqn:sym}
\min\{ \delta(A|B),\delta(B|A)\} - \tilde{I}(A,B) \leq \Delta_q - \Delta_c & \leq \Delta \delta(A|B).
\end{align}
Applying our assumption that $\delta(A|B) \leq \delta(B|A)$, and the observation that $\Delta_q - \Delta_c = I_q - I_c = \Delta I$, results in Eqn. \ref{eqn:symmetric_relation} as required.

\section{Example of Maximal Encodings}
In this section, we prove the assertion made in the paper that there always exists maximal encodings. Recall that we may define maximal encodings as follows:

\begin{defn}[Maximal Encoding]
Consider a bipartite quantum system with subsystems $A$ and $B$ that is described by density operator $\rho_{AB}$. The encoding of a random variable $\mathbf{K}$ that takes on values $k$ with probability $p_k$, by application of unitaries $U_k$ is a maximal encoding if and only if $I\left( \sum_k p_k U_k \; \rho_{AB} \; U_k^\dag \right) = 0\;$ for any $\rho_{AB}$, where $I(\rho_{AB})$ denotes the mutual information of $\rho_{AB}$.
\end{defn}

In particular, we prove the following.

\begin{thrm}\label{thm:encode}Suppose Alice's bipartition has dimension $d$, then $U_k$ is a maximum encoding whenever $\trho_{AB}$ is locally a maximally mixed state for any input state $\rho_A$ on Alice's bipartition.
\end{thrm}

\textbf{Proof:} To prove the result, it suffices to show that $\trho_{AB}$ is a product state. Consider an arbitrary projective measurement of the $B$ subsystem in some basis $\{\Pi_b\}$ on $\trho_{AB}$. Since these measurements commute with $U_k$, it follows that $\tr_B(\Pi_b \trho_{AB}) = \mathbf{I}/d$ for all $j$. Thus $\trho_{AB}$ must be a product state and the result follows. \qed

Therefore, any encoding that looks like a maximally mixing channel is a maximal encoding. One example, on a system of qubits, for example, is application of the set of unitary transformations $\{I,\sigma_x,\sigma_z,\sigma_x \sigma_z\}$. In an continuous variable mode with annihilation operator $a$, application of an operation selected uniformly from the set of displacement operators $\{D(\alpha) = \exp(\alpha a^\dag + \alpha^* a\}$ is also a maximal encoding.

\section{Application to Continuous Variables}
We specialize to the case where $A$ and $B$ are continuous variables modes, with respective quadrature operators $X_A,Y_A$ and $X_B,Y_B$ that obey the commutation relations $[X_j, Y_k] = 2i\delta_{jk}$. The resource state $\rho_{AB}$ created by displacement of two coherent vacuum states has a covariance matrix $\mathbf{\sigma}(\rho_{AB})$ of the form
\begin{equation}\label{eqn:idealmat}
\mathbf{\sigma}(\rho_{AB}) = \begin{pmatrix}
  V+1 & 0 & V & 0 \\
  0 & V+1 & 0 & -V \\
  V & 0 & V+1 & 0 \\
  0 & -V & 0 & V+1 \\
\end{pmatrix}
\end{equation}
where $V$ is the variance of the correlated noise added during the preparation of $\rho_{AB}$. The Gaussian discord, $\delta(A|B)$ shared between Alice and Bob is thus
\begin{equation}
\delta(A|B) = g\left(V+1\right) - 2g \left(\sqrt{2V + 1}\right)+ g\left(1 + \frac{2V}{2+V}\right).
\end{equation}
where $g(x) = x_+ \log_2 x_+ - \log_2 x_-$, and $x_\pm = \frac{1}{2}(x \pm 1)$~\cite{Giorda10}. Alice then encodes separate signals $x_s$ and $y_s$  governed respectively by Gaussian distributed random variables $\mathbf{X}_s$, $\mathbf{Y}_s$ of variance $V_s$ in the amplitude and phase of her mode by application of $\mathcal{E}_A(x_s,y_s) = \exp(-i x_s X_A/2)\exp(-i y_s Y_A/2)$. This results in a encoded state $\trho_{AB} = \int \mathcal{E}_A \rho_{AB} \mathcal{E}_A^\dag dx_s dy_s$ with covariance matrix
\begin{equation}\label{idealCVmat}
\mathbf{\sigma}(\trho_{AB}) = \begin{pmatrix}
  V+1 & 0 & V & 0 \\
  0 & V+1 & 0 & -V \\
  V & 0 & V+V_s+1 & 0 \\
  0 & -V & 0 & V+V_s+ 1 \\
\end{pmatrix}.
\end{equation}
Bob is tasked with extracting as much information about the encoded signal $({\bf X}_s,{\bf Y}_s)$. Should Bob be limited to a single measurement on each mode, his theoretical maximum performance is bounded above by
\begin{equation}
I_c = g\left(1 + \frac{2V}{V+2} + V_s\right) - g\left(1 + \frac{2V}{2+V}\right).
\end{equation}
by application of Eqn (\ref{eq:icreverse}). In contrast, if he also has the capacity to perform arbitrary coherent interactions, he's performance is given by
\begin{equation}\label{eqn:cv_q2_ideal}
I_q = S(\trho_{AB}) - S(\rho_{AB}) = g(\mu_+) + g(\mu_-) - 2 g(\sqrt{2V + 1}),
\end{equation}
and
\begin{equation}\nonumber
\mu_{\pm} = \sqrt{2V + 1 + \frac{V_s}{2}(V_s + 2V + 2 \pm \sqrt{(V_s+2)(4V +V_s + 2)})}.
\end{equation}
such that $\lim_{V_s \rightarrow \infty}(I_q - I_c) = \delta(\rho_{AB})$ as expected from Theorem \ref{thm:encode}.

The Holevo bound dictated by Eq. (\ref{eqn:cv_q2_ideal}) corresponds to an ideal theoretical protocol. In practical experiment, where we wish to demonstrate that coherent interaction leads to a definite advantage, such an ideal protocol is unnecessary. In our experiment, Bob coherently interacts his bipartite system via a 50-50 beamsplitter. The resulting beams are then measured in an appropriate quadrature basis. The resulting knowledge gained by Bob is given by
\begin{equation}
I_q^{\rm exp} = \log\left(1 + \frac{V_s}{2}\right)
\end{equation}
One can check that as $V \rightarrow \infty$, $I_q \rightarrow I_q^{\rm exp}$. This protocol is `almost optimal' provided the discord in the initial resource is large, and is thus sufficient for Bob to extract more knowledge that the incoherent limit $I_c$.

We experimentally prepare the aforementioned resource state $\rho_{AB}$, and encode within it the signals $({\bf X}_s,{\bf Y}_s)$. We then take on the role Bob, and attempt to measure some observable pairs $({\bf X}^{\rm exp}_s,{\bf Y}^{\rm exp}_s)$ such that $I({\bf X}_s,{\bf Y}_s;{\bf X}^{\rm exp}_s,{\bf Y}^{\rm exp}_s)$ is maximized. Theory dictates that when limited to a single local measurement on each subsystem, $I({\bf X}_s,{\bf Y}_s;{\bf X}^{\rm exp}_s,{\bf Y}^{\rm exp}_s) \leq I_c$. Thus, our experimental conclusively demonstrates that the capacity for coherent processing can harness uniquely quantum correlations provided the above inequality is violated. The magnitude in which we can violate this inequality
\begin{equation}
\Delta I^{\rm exp}=I({\bf X}_s,{\bf Y}_s;{\bf X}^{\rm exp}_s,{\bf Y}^{\rm exp}_s) - I_c \def \delta^{\rm exp}(\rho_{AB})
\end{equation}
then defines the observed discorded assisted quantum advantage.

We can also compare this rate to the optimal known decoding scheme when limited a single local measure of $A$ and $B$. In this scheme, Bob makes simultaneous quadrature
measurement of the two beams. 
The information Bob can extract from a state having covariance matrix
in Eq.~\ref{idealCVmat} is
\begin{equation}
I_c^{prot} = \log\left(1+\frac{1+V}{1+2V}V_s \right)
\end{equation}

\section{Experimental Details}
In this section, we discuss the details of the experiment carried out including the hardware, implementation, processing and the sources of errors.

The experimental setup is shown in Fig.~\ref{fig1}. The entire experiment uses a single 1064 nm Nd:YAG laser source. The light is passed through a mode cleaner cavity to provide a broadly shot-noise limited coherent light source from 0.9 MHz.

A small portion of the original light power is split to provide the two modes of the bipartite state, each passed through a pair of phase and amplitude electro-optic modulators (EOM). The laser source also provides a bright field as a local oscillator for homodyne detection. The homodyne efficiency of the set-up is estimated at 91\%. This is limited by the quantum efficiency of the diodes (estimated at 93\%) and typical mode-matching qualities of 98\%, generally limited by the mode distortions introduced by the EOM's. The detectors are electronically matched to provide a common mode rejection of 45 db.

For the purposes of control of the measured quadrature, appropriate phase and amplitude modulation are introduced onto the beam. This allows us to verify the suppression of noise contributions from parasitic modulations, and its contribution to error in discerning the quadrature. Typical suppression of the orthogonal quadrature is greater than 25 db.

The resource state is created by displacing both modes A and B by the same magnitude both in amplitude and phase quadratures. The displacement of the phase quadrature and amplitude quadrature of modes A and B is chosen to be correlated and anti-correlated respectively. Each white noise signal used is generated from a single function generator that provides a broadband white noise signal up to 10 MHz. An electronic gain and delay is then introduced on mode B to synchronise it to mode A at 3.6 MHz. Magnitudes for the white noise encoded on phase and amplitude quadratures are matched to ensure the closest realisation to the ideal symmetric covariance matrix (see Equation~\ref{idealCVmat}).

Two additional function generators that generate broadband Gaussian white noise up to 10 MHz are used to provide the signal encoding for the amplitude and phase quadratures. The electronic resource is split, with one portion recorded, whilst the remainder is sent to the existing EOM for displacement of mode A. The encoded signal variance of the phase quadrature is electronically attenuated to match the variance of the amplitude quadrature.

In the setup where coherent interactions between the subsystems are not permitted, the resource state is characterised by first measuring the amplitude quadrature of both beams and subsequently measuring the phase quadrature of both beams. For the coherent processing case, the two beams are interfered in phase on a 50:50 beam splitter. The amplitude quadrature of the bright output and the phase quadrature of the dark output are sampled using the same homodyne detectors.

For each separate homodyne detection $10^6$ data points are  sampled at 20 Msamp per second using a digital acquisition system. The process is repeated five times for each data point to provide sufficient statistics. Each of the five data set is divided into two sets. These data is then digitally filtered to 3.6-3.8 MHz and then re-sampled.

\section{Model}
Whilst ideally, we strive to achieve the state described by the covariance matrix in Eq.~\ref{idealCVmat}, in practise the actual state is never such. To refine the experimental model, we include effects of imperfect correlations, non linear transmission and modulation losses, excess noise and unbalanced beam splitter ratio.
The covariance matrix of the bipartite state is a function of the input signal, input noise and the quantum noise. It can be written as $C_0=\hat{v}^\dagger \hat{v}$ where
\begin{equation}
\hat{v} = \left( \vec{A}_X, \vec{A}_Y,
\vec{B}_X, \vec{B}_Y \right)
\end{equation}
and $\vec{A}_{X(Y)}$ and $\vec{B}_{X(Y)}$ are the modulations on the amplitude (phase) quadratures on beams A and B written as a linear combination of eight independent inputs: the input signals for $X(Y)$, $\sigma_{sx(y)}$, the input classical noise for $X(Y)$, $\sigma_{nx(y)}$ and the vacuum noises in $X(Y)$ in beam A and beam B, $\sigma_v$. We write $\hat{v}$ as
\begin{equation}
\hat{v} = \begin{pmatrix}
\eta^A_{XX} \sigma_{sx}&\eta^A_{YX} \sigma_{sx}&0&0\\
\eta^A_{XY} \sigma_{sy}&\eta^A_{YY} \sigma_{sy}&0&0\\
\beta^A_{XX} \sigma_{nx}&\beta^A_{YX} \sigma_{nx}&\beta^B_{XX} \sigma_{nx}&\beta^B_{YX} \sigma_{nx}\\
\beta^A_{XY} \sigma_{ny}&\beta^A_{YY} \sigma_{ny}&\beta^B_{XY} \sigma_{ny}&-\beta^B_{YY} \sigma_{ny}\\
\xi^A_X \sigma_{v}&0&0&0\\
0&\xi^A_Y \sigma_{v}&0&0\\
0&0&\xi^B_X \sigma_{v}&0\\
0&0&0&\xi^B_Y \sigma_{v}
\end{pmatrix}.
\end{equation}
The coefficients $\eta$ and $\beta$ characterise the linear correlations between the quadrature modulations and the applied signal and noise voltages. The terms $\eta_{XY}$ and $\eta_{YX}$ represent the parasitic cross correlations which can be due to imperfect modulation quadrature. Ideally we want these to be zero. A non zero correlation will degrade the mutual information. The terms $\eta_{XX}$ and $\eta_{YY}$ are the correlations between the signal and the quadrature modulation. Imperfect correlation will again degrade both the resulting information for both the coherent and incoherent interaction. We also need the noise on both beams to have the same magnitudes so that they cancel each other at the beam splitter performing the coherent interaction. Two other requirements to observe the maximum quantum advantage are for the noises and signals in both quadratures to have the same magnitudes. If they are not equal, the penalty incurred when doing a classical measurement will be less than one unit of shot noise. Finally, the coefficients $\xi$ characterise excess noises in the quadratures.  For the coherent case, we also include a small nonlinear loss that increases with the signal variance around the order of $\eta_{loss}=0.0001 \sigma_{sx(sy)}^2+ 0.00003 \sigma_{sx(sy)}^4$ just before the beam splitter. This is attributed to the nonlinear response of the electro-optic modulators and gives rise to the observed plateauing of the quantum advantage in Fig.~\ref{fig3}. The loss is simulated by propagating the covariance matrix $C_0$ through a beam splitter and tracing over the output of the vacuum port to get the new covariance matrix:
\begin{equation}
C_1^A = \tr_v \{BS(\eta_{loss})\cdot C_0^A\oplus C_v \cdot BS(\eta_{loss})^\dagger \}
\end{equation}
where
\begin{equation}
C_v =
\begin{pmatrix}
\sigma_v^2&0\\
0& \sigma_v^2
\end{pmatrix}
\end{equation}
is the covariance matrix for the vacuum input and
\begin{equation}
BS(\eta) =
\begin{pmatrix}
\sqrt \eta&0 & -\sqrt{1-\eta}&0\\
0&\sqrt \eta&0 & -\sqrt{1-\eta}\\
\sqrt{1- \eta}&0 & \sqrt{\eta}&0\\
0&\sqrt{1- \eta}&0 & \sqrt{\eta}\\
\end{pmatrix}
\end{equation}
is the beam splitter transformation with transmission $\eta$. $C_0^A = \tr_B\{C_0\}$ is the covariance matrix for beam $A$.  In the case where coherent interactions are permitted, the beams A and B are then propagated through an interference beam splitter with transmission coefficient $\eta_{i}=0.48$ and the relative phase between the two beams fixed at $\phi_A-\phi_B=0$. The output covariance matrix is then
\begin{equation}
C_2 = BS(\eta_{i}) PS(\phi_A,\phi_B)\cdot C_1 \cdot PS(\phi_A,\phi_B)^\dagger BS(\eta_{i})^\dagger
\end{equation}
where $PS(\phi_A,\phi_B)$ = $PS(\phi_A)\oplus PS(\phi_B)$ shifts the
phases of beam $A(B)$ by $\phi_{A(B)}$ with
\begin{equation}
PS(\phi) =
\begin{pmatrix}
\cos \phi& -\sin \phi\\
\sin \phi& \cos \phi
\end{pmatrix}\;.
\end{equation}
The homodyne efficiencies are modeled as a vacuum noise contaminating the signal. Moreover, we take into account an imperfect locking angle between the local oscillator and the signal modeled as a rotation of the beam quadrature before the measurement
\begin{equation}
C_3^{A} = \tr_v \{BS(\eta_{lo}^{A}) PS(\phi_{lo}^{A})\cdot C_2^{A} \cdot PS(\phi_{lo}^{A})^\dagger BS(\eta_{lo}^A)^\dagger \}
\end{equation}
and a similar expression for $C_3^B$ with $\phi_{lo}^A = 0$ and $\phi_{lo}^B = \pi/2$. Finally, tracing over the phase quadrature gives the measured output of the detectors in the coherent interaction setup $SX_{measured}=\tr_{Y} \{C_3^A\}$ and $SY_{measured} = \tr_{Y} \{C_3^B\}$. In the incoherent interaction case, the covariance matrix $C_0$ is directly propagated through to the homodyne detection to sequentially measure both the $X$ and $Y$ quadratures of both beams. The information rate without coherent interactions are then calculated using the full covariance matrix,.

\end{appendix}

\end{document}